\documentclass[conference]{IEEEtran}
\normalsize
\ifCLASSINFOpdf
\else
\fi
\usepackage{array}
\usepackage{color}
\usepackage{amsfonts}
\usepackage{amssymb}
\usepackage{amsmath}
\usepackage[dvips]{graphicx}
\usepackage{cite}
\usepackage{balance}
\usepackage{caption}
\usepackage{subcaption}
\usepackage{textcomp}
\usepackage{gensymb}
\usepackage{float}
\usepackage{tabularx}
\usepackage{multirow}
\usepackage{amsthm}
\newtheorem{theorem}{Theorem}
\newtheorem{corollary}{Corollary}

\begin{document}

\title{On the Secrecy Capacity of Fisher-Snedecor \\ $\mathcal{F}$
Fading Channels}
{\color{red} \large{This paper has been submitted to GlobeCom 2018}}
\author{Osamah~S.~Badarneh$^{1}$,~Paschalis C. Sofotasios$^{2, 3}$,~Sami~Muhaidat$^{2, 4}$,~Simon~L.~Cotton$^{5}$, \\ Khaled~Rabie$^{6}$,~and~Naofal~Al-Dhahir$^{7}$ \\ \\
$^{1}$Department of Electrical Engineering, University of Tabuk, P. O. Box 741, Tabuk, KSA   \\   (e-mail: $\rm obadarneh@ut.edu.sa$) \\
$^{2}$Department of Electrical and Computer Engineering, Khalifa University of Science and Technology,  \\ P.  O.  Box   127788, Abu Dhabi, UAE  \,    (e-mail: $\rm \{ paschalis.sofotasios; sami.muhaidat \}@ku.ac.ae$) \\
$^{3}$Department of Electronics and Communications Engineering, Tampere University of Technology, \\ FI-33101, Tampere, Finland \, (e-mail: $\rm  paschalis.sofotasios@tut.fi$)  \\
$^{4}$Institute for Communication Systems, University of Surrey, GU2 7XH, Guildford, UK\\   (e-mail:
$\rm sami.muhaidat@surrey.ac.uk$)  \\
$^{5}$Institute of Electronics, Communications and Information Technology, Queen's University Belfast,   \\ BT3 9DT,  Belfast, UK  (e-mail:
$\rm   simon.cotton@qub.ac.uk$)  \\
$^{6}$School of Electrical Engineering, Manchester Metropolitan University, M15 6BH, \\ Manchester, UK    \, (e-mail: $\rm  k.rabie@mmu.ac.uk$)   \\
$^{7}$Department of Electrical \& Computer Engineering, University of Texas at Dallas, \\ Richardson, TX 75080, USA, \, \, (e-mail: $ \rm aldhahir@utdallas.edu$)}
\maketitle	
\begin{abstract}
The performance of physical-layer security of the classic Wyner's wiretap model over Fisher-Snedecor $\mathcal{F}$ composite fading channels is considered in this work. Specifically, the main channel (i.e., between the source and the legitimate destination) and the eavesdropper's channel (i.e., between the source and the illegitimate destination) are assumed to experience independent quasi-static Fisher-Snedecor $\mathcal{F}$ fading conditions, which have been shown to be encountered in realistic wireless transmission scenarios in conventional and emerging communication systems. In this context, exact closed-form expressions for the average secrecy capacity (ASC) and the probability of non-zero secrecy capacity (PNSC) are derived. Additionally, an asymptotic analytical expression for the ASC is also presented. The impact of shadowing and multipath fading on the secrecy performance is investigated. Our results show that increasing the fading parameter of the main channel and/or the shadowing parameter of the eavesdropper's channel improves the secrecy performance. The analytical results are compared with Monte-Carlo simulations to validate the analysis.
\end{abstract}
\IEEEpeerreviewmaketitle
\section{Introduction}
\IEEEPARstart{S}{hannon} introduced in his seminal work \cite{shannon} the concept of information-theoretic secrecy by investigating the transmission of a secret message between a source, a legitimate receiver and an eavesdropper where the source and the legitimate receiver share a secret key. In order to achieve perfect secrecy, the key rate should be at least as large as the message rate. Later in \cite{Wyner}, Wyner introduced the notion of the wiretap channel that enables a source to send a secret message to a legitimate receiver in the presence of an eavesdropper without using a shared key such that almost no information about the secret message is leaked to the eavesdropper.

In the literature, the performance of physical-layer security (PLS) has been analyzed using several metrics in the context of different wireless communication systems and over various fading channels \cite{JMRomero,WLiu,MBenammar,HLei3,HLei1,HLei2,Bhargav,Bhargav2,XLiu,XianLiu,XLiu1,ABoulogeorgos,GPan}. Examples of these performance metrics include the average secrecy capacity (ASC), secure outage probability (SOP), and the probability of strictly positive secrecy capacity (SPSC). In this context, a general and compact analysis framework for the wireless information-theoretic security of arbitrarily-distributed fading channels was presented in \cite{JMRomero}. In \cite{WLiu}, the authors investigated PLS in a random wireless network where both legitimate and eavesdropping nodes are randomly deployed. The secrecy capacity of the wiretap broadcast channel with an external eavesdropper was studied in \cite{MBenammar}. The authors in \cite{HLei3} studied the ASC performance of the classic Wyner's model over $\alpha$-$\mu$ fading channels, while the authors in \cite{HLei1} derived closed-form expressions for the ASC, SOP, and the SPSC assuming a classic Wyner's wiretap model over generalized-$K$ fading channels.

In the same context, the SOP and SPSC performance over generalized gamma fading channels was analyzed in \cite{HLei2}. Novel analytical expressions for the SPSC and a lower bound on SOP were derived over independent and non-identically distributed $\kappa$-$\mu$ fading channels in \cite{Bhargav}. The derived expressions were then used to study the performance of different emerging wireless applications, such as cellular device-to-device, peer-to-peer, vehicle-to-vehicle and body centric communications. In \cite{Bhargav2}, the same authors derived analytical expressions for the SPSC and a lower bound on SOP over $\alpha$-$\mu$/$\kappa$-$\mu$ and $\kappa$-$\mu$/$\alpha$-$\mu$ fading channels. The SPSC performance for wireless communication systems under Rician, log-normal, and Weibull fading was analyzed in \cite{XLiu,XianLiu,XLiu1}. The effect of in-phase and quadrature imbalance on the ASC was studied in \cite{ABoulogeorgos}, whereas the performances of secure communications, in terms of the ASC, SOP, and probability of non-zero secrecy capacity (PNSC), over independent and correlated lognormal shadowing channels as well as composite fading channels were investigated in \cite{GPan}.
\begin{figure}[t] \centering
         \includegraphics[width=0.6\columnwidth]{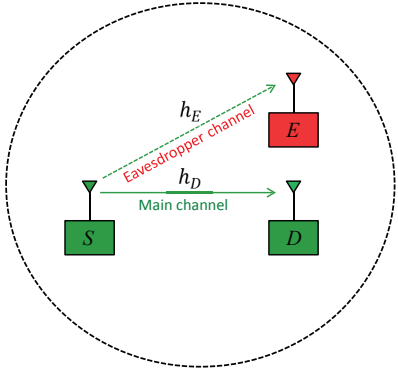}%
       \caption{Example of a single-hop wireless system with an eavesdropper.}\label{sysmod}
\end{figure}

The studies above have shown that the PLS performance depends heavily upon the fading characteristics of the channels between the communicating nodes as well as the eavesdroppers. Recently, an accurate and tractable composite fading model was proposed to characterize the combined effects of multipath fading and shadowing. The so-called Fisher-Snedecor $\mathcal{F}$ distribution was introduced in \cite{Fisher} under the assumption that the scattered multipath follows a Nakagami-$m$ distribution, while the root-mean-square signal is shaped by an inverse Nakagami-$m$ random variable. It is worth remarking that owing to its generality, several well-known fading distributions such as the Nakagami-$m$ and Rayleigh distributions can be obtained as special cases. In addition, the Fisher-Snedecor $\mathcal{F}$ distribution has been shown, both theoretically and experimentally, to provide a good approximation to other composite fading distributions such as the generalized-$K$ model with lower computational complexity \cite{Fisher}.

To the best of the authors' knowledge, the PLS performance over Fisher-Snedecor $\mathcal{F}$ composite fading has not been investigated in the open literature. In this paper, we consider a wireless system in which the main channel and the eavesdropper's channel experience independent quasi-static Fisher-Snedecor $\mathcal{F}$ fading, which is a practically realistic assumption. In particular, the ASC and PNSC of the classic Wyner's model over Fisher-Snedecor $\mathcal{F}$ composite fading channels are derived in closed-form. Additionally, an asymptotic ASC analysis in a high signal-to-noise ratio (SNR) regime (i.e., when the destination node is located close to the source node) is also provided.

The rest of this paper is organized as follows. In the next section, we present the system and channel models, whereas in Section \ref{sec3}, we derive the ASC and PNSC expressions. Then, we quantify the impact of multipath fading and shadowing on the secrecy performance in Section \ref{sec6}, while some concluding remarks are provided in Section \ref{sec7}.

\section{System and Channel Models}
\begin{figure}[t] \centering
         \includegraphics[width=0.8\columnwidth]{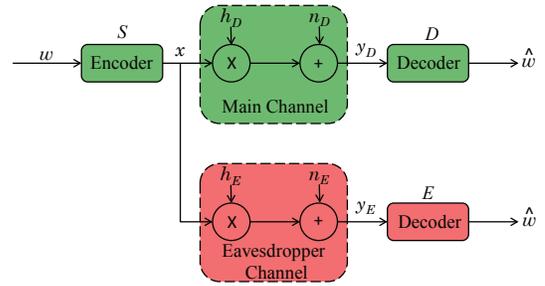}%
       \caption{The system model of the wiretap channel in the presence of Fisher-Snedecor $\mathcal{F}$ fading.}\label{sysmod1}
\end{figure}
We consider a wiretap channel \cite{Wyner} which consists of a legitimate source ($S$), a legitimate destination ($D$), and an eavesdropper ($E$). In this system, the legitimate source $S$ sends a confidential message to the legitimate destination $D$ over the main channel while the eavesdropper $E$ attempts to decode this message from its received signal, as depicted in Fig. \ref{sysmod}. The main channel ($S$$\rightarrow$$D$) and the eavesdropper's channel ($S$$\rightarrow$$E$) experience independent quasi-static Fisher-Snedecor $\mathcal{F}$ fading \cite{Fisher}. The fading coefficients, for both links, remain constant during a transmission block time but vary independently from one block to another. Also, the channel state information (CSI) is assumed to be known at both the transmitter and the destination. Additionally, we assume that the noise at all nodes is zero-mean and unit-variance additive white Gaussian noise (AWGN).

The system model of the wiretap channel in the presence of Fisher-Snedecor $\mathcal{F}$ fading is shown in Fig. \ref{sysmod1}. Based on this, the signals received at $D$ and $E$ can be expressed, respectively, as
\begin{equation}\label{recyd}
  y_{D} = \sqrt{P_t}h_D x + n_D,
\end{equation}
and
 \begin{equation}\label{recye}
  y_{E} = \sqrt{P_t}h_E x + n_E,
\end{equation}
where $P_t$ denotes the transmitted power at the source $S$, $h_D$ and $h_E$ are the channel coefficients of the main and eavesdropper's channels, respectively, $x$ is the normalized transmitted signal (i.e., $\mathbb{E}[|x|^{2}]=1$, whereby $\mathbb{E}[\cdot]$ denotes expectation operator), and $n_D$ and $n_E$ are the AWGN at $D$ and $E$ with variances of $\sigma_{D}^{2}$ and $\sigma_{E}^{2}$, respectively. The instantaneous output SNRs at $D$ and $E$ are, respectively, given by $\gamma_{D} = \overline{\gamma}_{D}\left\vert h_D\right\vert^2$ and $\gamma_{E} = \overline{\gamma}_{E}\left\vert h_E\right\vert^2$,
where $\overline{\gamma}_{D}=P_t/\sigma_{D}^{2}$ and $\overline{\gamma}_{E}=P_t/\sigma_{E}^{2}$ represent the average SNR of the main and eavesdropper links, respectively.

It is recalled that the main and eavesdropper links experience independent but not necessarily identically distributed (i.n.i.d) Fisher-Snedecor $\mathcal{F}$ fading. Hence, the probability density function (PDF) of the instantaneous SNR, $\gamma_{k}$, of a Fisher-Snedecor $\mathcal{F}$ fading envelope can be expressed as \cite{Fisher}
\begin{align}\label{pdfr}
f_{k}(\gamma_{k})={{a_{k}^{m_{k}}}\over B(m_{k},m_{s_{k}})}{\gamma_{k}^{m_{k}-1}\over\left(1+a_{k}\gamma_{k}\right)^{m_{k}+m_{s_{k}}}},
\end{align}
which can also be rewritten using \cite[Eq. (8.4.2.5)]{a:pru} and \cite[Eq. (8.2.2.15)]{a:pru} as follows
\begin{align}\label{pdfr1}
f_{k}(\gamma_{k})={1\over\Gamma(m_{k})\Gamma(m_{s_{k}})}\gamma^{-1}_{k}\mathrm{G}_{1,1}^{1,1}\left(a_k\gamma_k\left\vert \begin{matrix} 1-m_{s_{k}}\\ m_{k}\end{matrix}\right.\right),
\end{align}
where $a_{k}={m_{k}\over m_{s_{k}}\overline{\gamma}_{k}}$, $m$ and $m_{s}$ are the multipath fading severity and shadowing parameters, respectively, $\overline{\gamma}=\mathbb{E}[\gamma]$ is the mean power, $B(\cdot,\cdot)$ is the beta function \cite[Eq. (8.384.1)]{i:ryz}, and $\mathrm{G}_{u,v}^{s,t}(\cdot)$ is the Meijer's G-function \cite[Eq. (8.3.1.1)]{a:pru}.

The corresponding cumulative distribution function (CDF) of $\gamma_{k}$ can be found, using \cite[Eq. (26)]{Adamchik}, to be
\begin{align}\label{cdfr}
F_{k}(\gamma_{k})={1\over\Gamma(m_{k})\Gamma(m_{s_{k}})}\mathrm{G}_{2,2}^{1,2}\left(a_k\gamma_k\left\vert \begin{matrix} 1-m_{s_{k}},1\\ m_{k},0\end{matrix}\right.\right).
\end{align}
Note that the Fisher-Snedecor $\mathcal{F}$ distribution includes other well-known fading distributions, namely, the Nakagami-$m$ distribution when $m_{s}\rightarrow\infty$, Rayleigh distribution when $m_{s}\rightarrow\infty$ and $m=1$ and one-sided Gaussian distribution when $m_{s}\rightarrow\infty$ and $m={1/2}$.

\section{Secrecy Capacity Analysis}\label{sec3}
In this analysis, we consider an active eavesdropping scenario where the CSI of the eavesdropper's channel is known at the source $S$. As such, $S$ can adapt the achievable secrecy rate $R$ such that $R\leq R_{s}$. The maximum achievable secrecy rate $C_s=R_s$ can be characterized as \cite{Bloch}
\begin{equation}\label{defsec}
{C_s} = \{[C_{D} -  C_{E}]^+ \},
\end{equation}
where $C_D =\ln\left({1 + {\gamma _D}}\right)$ and $C_E = \ln\left({1 + {\gamma _E}}\right)$ are the capacities of the AWGN channels used by the main and eavesdropper channels, respectively, and $[z]^+=\max\{z,0\}$. Thus, (\ref{defsec}) can be expressed as
\begin{align}\label{defsec1}
C_{s} = \begin{cases} \ln[1+\gamma_{D}]-\ln[1+\gamma_{E}], & \gamma_{D}>\gamma_{E}\\ 0, & \gamma_{D}\leq\gamma_{E} \end{cases}
\end{align}
\subsection{Existence of Secrecy Capacity}
\begin{theorem}[The Probability of Existence of a Non-zero Secrecy Capacity] The PNSC for the system under consideration when the main channel and eavesdropper's channel experience i.n.i.d. Fisher-Snedecor $\mathcal{F}$ fading is given by
\begin{align}\label{sxcs1}
&C_{s}^{\rm{Exist}} = {1\over\Gamma(m_{D})\Gamma(m_{s_{D}})\Gamma(m_{E})\Gamma(m_{s_{E}})}\cr&\qquad\qquad\qquad\times\mathrm{G}_{3,3}^{3,2}\left({a_D\over a_E}\left\vert \begin{matrix} 1-m_{s_{D}},1-m_E,1\\ m_D,m_{s_{E}},0\end{matrix}\right.\right).&
\end{align}
\end{theorem}

\begin{IEEEproof}
It follows from (\ref{defsec1}) that the secrecy capacity is positive when $\gamma_{D}>\gamma_E$ and is zero when $\gamma_{D}\leqslant\gamma_E$. Knowing that both the main channel and the eavesdropper's channel experience independent fading, the probability of existence of a non-zero secrecy capacity can be written as
\begin{align}\begin{aligned}
C_{s}^{\rm{Exist}}&=\Pr\left\{ {C _s> 0} \right\}\\&=\Pr\left\{ {\gamma _D> \gamma _E} \right\}\\& = \int\limits_{0}^{\infty}f_{D}(\gamma_D)\int\limits_{0}^{\gamma _D}f_{E}(\gamma_E)d{\gamma_E}d{\gamma_D}.\label{xcs}\end{aligned}
\end{align}
Capitalizing on the definition $F_X(x)\triangleq\int_{0}^{x}f_X(y)dy$, equation (\ref{xcs}) can be rewritten as
\begin{align}\begin{aligned}
C_{s}^{\rm{Exist}}= \int\limits_{0}^{\infty}f_{D}(\gamma_D)F_E(\gamma_D)d{\gamma_D}.\label{xcs1}\end{aligned}
\end{align}
Therefore, by making the necessary change of variables in (\ref{pdfr1}) and (\ref{cdfr}), equation (\ref{xcs}) becomes
\begin{align}\label{xcs1}
&C_{s}^{\rm{Exist}} = {1\over\Gamma(m_{D})\Gamma(m_{s_{D}})\Gamma(m_{E})\Gamma(m_{s_{E}})}\int\limits_0^\infty \gamma_{D}^{-1} \cr\times&\mathrm{G}_{1,1}^{1,1}\left(a_D\gamma_D\left\vert \begin{matrix} 1-m_{s_{D}}\\ m_{D}\end{matrix}\right.\right)\mathrm{G}_{2,2}^{1,2}\left(a_E\gamma_D\left\vert \begin{matrix} 1-m_{s_{E}},1\\ m_{E},0\end{matrix}\right.\right)d{\gamma_D},\cr&
\end{align}
The above integral can be obtained in closed-form using \cite[Eq. (2.24.1.1)]{a:pru}. Thus, after the necessary change of variables, equation (\ref{sxcs1}) is obtained, which completes the proof.
\end{IEEEproof}
\subsection{Average Secrecy Capacity Analysis}
\begin{theorem}[Average Secrecy Capacity] The ASC for the system under consideration when the main channel and eavesdropper's channel experience i.n.i.d. Fisher-Snedecor $\mathcal{F}$ fading is given by (\ref{sasc}), at the top of the next page, where $\mathrm{G}_{p_1,q_1:p_2,q_2:p_3,q_3}^{m_1,n_1:m_2,n_2:m_3,n_3}(\cdot)$ is the bivariate Meijer's G-function \cite[II. 13]{THNguyen}.
\begin{figure*}
\begin{align}\begin{aligned}
&\overline{C}_{s}^{\rm{ASC}}={\mathrm{G}_{2,2:1,1:2,2}^{2,1:1,1:1,2}\left(\begin{array}{c} a_D,a_E\end{array}\left\vert \begin{array}{c} 0,1 \\ 0,0 \end{array}\right.\left\vert \begin{array}{c} 1-m_{s_{D}}\\m_{D} \end{array}\right\vert \begin{array}{c}  1-m_{s_{E}},1\\m_{E},0 \end{array}\right)\over\Gamma(m_{D})\Gamma(m_{s_{D}})\Gamma(m_{E})\Gamma(m_{s_{E}})}\cr&\qquad\qquad\qquad+
{\mathrm{G}_{2,2:1,1:2,2}^{2,1:1,1:1,2}\left(\begin{array}{c} a_E,a_D\end{array}\left\vert \begin{array}{c} 0,1 \\ 0,0 \end{array}\right.\left\vert \begin{array}{c} 1-m_{s_{E}}\\m_{E} \end{array}\right\vert \begin{array}{c}  1-m_{s_{D}},1\\m_{D},0 \end{array}\right)\over\Gamma(m_{D})\Gamma(m_{s_{D}})\Gamma(m_{E})\Gamma(m_{s_{E}})}
-{\mathrm{G}_{3,3}^{3,2}\left(a_E\left\vert \begin{matrix} 1-m_{s_{E}},0,1\\ m_{E},0,0\end{matrix}\right.\right)\over\Gamma(m_{E})\Gamma(m_{s_{E}})}.&
\label{sasc}\end{aligned}\end{align}
\hrulefill
\end{figure*}
\begin{figure*}
\begin{align}\begin{aligned}
 & \overline{C}_{s}^{\rm{ASC, asy}}={a_{D}^{m_{D}}\mathrm{G}_{4,4}^{3,3}\left(a_E\left\vert \begin{matrix} 1-m_{s_{E}},1,-m_D,1-m_D\\ m_{E},-m_D,-m_D,0\end{matrix}\right.\right)\over B(m_D,m_{s_{D}})\Gamma(m_{E})\Gamma(m_{s_{E}})}\cr&\qquad\qquad\qquad+
{\mathrm{G}_{2,2:1,1:2,2}^{2,1:1,1:1,2}\left(\begin{array}{c} a_E,a_D\end{array}\left\vert \begin{array}{c} 0,1 \\ 0,0 \end{array}\right.\left\vert \begin{array}{c} 1-m_{s_{E}}\\m_{E} \end{array}\right\vert \begin{array}{c}  1-m_{s_{D}},1\\m_{D},0 \end{array}\right)\over\Gamma(m_{D})\Gamma(m_{s_{D}})\Gamma(m_{E})\Gamma(m_{s_{E}})}
-{\mathrm{G}_{3,3}^{3,2}\left(a_E\left\vert \begin{matrix} 1-m_{s_{E}},0,1\\ m_{E},0,0\end{matrix}\right.\right)\over\Gamma(m_{E})\Gamma(m_{s_{E}})}.&
\label{asymasc}\end{aligned}\end{align}
\hrulefill
\end{figure*}
\end{theorem}
\begin{IEEEproof}
Invoking independence between the main channel and the eavesdropper's channels, the ASC is given by
\begin{align}\begin{aligned} \overline{C}_{s}^{\rm{ASC}} &= \mathbb{E}\left[{{C_s}\left({{\gamma _D},{\gamma _E}} \right)} \right]\\& = \int\limits_0^\infty {\int\limits_0^\infty {{C_s}\left({{\gamma _D},{\gamma _E}} \right)f\left({{\gamma _D},{\gamma _E}} \right)d{\gamma _D}d{\gamma _E}} }\\ & = {\mathcal{J}_1} + {\mathcal{J}_2} - {\mathcal{J}_3},\label{defasc}\end{aligned}\end{align}
where $f\left({{\gamma _D},{\gamma _E}} \right)$ is the joint PDF of $\gamma _D$ and $\gamma _E$, and
\begin{align}\label{cs1}
{\mathcal{J}_1} = \int\limits_0^\infty {\ln \left({1 + {\gamma _D}} \right){f_D}\left({{\gamma _D}} \right){F_E}\left({{\gamma _D}} \right)d{\gamma _D}},
\end{align}
\begin{align}\label{cs2}
{\mathcal{J}_2} = \int\limits_0^\infty {\ln \left({1 + {\gamma _E}} \right){f_E}\left({{\gamma _E}} \right){F_D}\left({{\gamma _E}} \right)d{\gamma _E}},
\end{align}
\begin{align}\label{cs3}
&{\mathcal{J}_3} = \int\limits_0^\infty {\ln \left({1 + {\gamma _E}} \right){f_E}\left({{\gamma _E}} \right)d{\gamma _E}}.&
\end{align}

By making the necessary change of variables in (\ref{pdfr1}) and (\ref{cdfr}), then $\mathcal{J}_1$ becomes
\begin{align}\label{j1}
&{\mathcal{J}_1} = \int\limits_0^\infty{\gamma_{D}^{-1} {\ln \left({1 + {\gamma _D}} \right)}\over\Gamma(m_{D})\Gamma(m_{s_{D}})\Gamma(m_{E})\Gamma(m_{s_{E}})}\cr \times&\mathrm{G}_{1,1}^{1,1}\left(a_D\gamma_D\left\vert \begin{matrix} 1-m_{s_{D}}\\ m_{D}\end{matrix}\right.\right)\mathrm{G}_{2,2}^{1,2}\left(a_E\gamma_E\left\vert \begin{matrix} 1-m_{s_{E}},1\\ m_{E},0\end{matrix}\right.\right)d{\gamma_D}.\cr&
\end{align}
Representing the logarithmic function in terms of Meijer's G-function \cite[Eq. (3.4.6.5)]{a:pru}, then (\ref{j1}) becomes
\begin{align}\label{j11}
&{\mathcal{J}_1} = \int\limits_0^\infty {\gamma_{D}^{-1}\mathrm{G}_{2,2}^{1,2}\left(\gamma_D\left\vert \begin{matrix} 1, 1\\ 1, 0\end{matrix}\right.\right)\over\Gamma(m_{D})\Gamma(m_{s_{D}})\Gamma(m_{E})\Gamma(m_{s_{E}})}\cr \times&\mathrm{G}_{1,1}^{1,1}\left(a_D\gamma_D\left\vert \begin{matrix} 1-m_{s_{D}}\\ m_{D}\end{matrix}\right.\right)\mathrm{G}_{2,2}^{1,2}\left(a_E\gamma_D\left\vert \begin{matrix} 1-m_{s_{E}},1\\ m_{E},0\end{matrix}\right.\right)d{\gamma_D}.\cr&
\end{align}
With the help of \cite{Wolfram}, ${\mathcal{J}_1}$ can be obtained in closed-form as
\begin{align}\label{sj1}
 &{\mathcal{J}_1} ={1\over\Gamma(m_{D})\Gamma(m_{s_{D}})\Gamma(m_{E})\Gamma(m_{s_{E}})}\times\cr&
\mathrm{G}_{2,2:1,1:2,2}^{2,1:1,1:1,2}\left(\begin{array}{c} a_D,a_E\end{array}\left\vert \begin{array}{c} 0,1 \\ 0,0 \end{array}\right.\left\vert \begin{array}{c} 1-m_{s_{D}}\\m_{D} \end{array}\right\vert \begin{array}{c}  1-m_{s_{E}},1\\m_{E},0 \end{array}\right),\cr&
\end{align}
whereas, similarly, the integral ${\mathcal{J}_2}$ can be obtained as
\begin{align}\label{sj2}
 &{\mathcal{J}_2} ={1\over\Gamma(m_{D})\Gamma(m_{s_{D}})\Gamma(m_{E})\Gamma(m_{s_{E}})}\times\cr&
\mathrm{G}_{2,2:1,1:2,2}^{2,1:1,1:1,2}\left(\begin{array}{c} a_E,a_D\end{array}\left\vert \begin{array}{c} 0,1 \\ 0,0 \end{array}\right.\left\vert \begin{array}{c} 1-m_{s_{E}}\\m_{E} \end{array}\right\vert \begin{array}{c}  1-m_{s_{D}},1\\m_{D},0 \end{array}\right).\cr&
\end{align}

Furthermore, the integral ${\mathcal{J}_3}$ can be rewritten using (\ref{pdfr1}) and (\ref{cs3}) as follows
\begin{align}\label{j3}
& {\mathcal{J}_3}={1\over\Gamma(m_{E})\Gamma(m_{s_{E}})}\int\limits_0^\infty \gamma_{E}^{-1}\mathrm{G}_{2,2}^{1,2}\left(\gamma_E\left\vert \begin{matrix} 1, 1\\ 1, 0\end{matrix}\right.\right)\cr &\qquad\qquad\qquad\qquad\times\mathrm{G}_{1,1}^{1,1}\left(a_E\gamma_E\left\vert \begin{matrix} 1-m_{s_{E}}\\ m_{E}\end{matrix}\right.\right)d{\gamma_E},&
\end{align}
which can be solved with the help of \cite[Eq. (2.24.1.1)]{a:pru} as
\begin{align}\label{sj3}
 {\mathcal{J}_3}={1\over\Gamma(m_{E})\Gamma(m_{s_{E}})}\mathrm{G}_{3,3}^{3,2}\left(a_E\left\vert \begin{matrix} 1-m_{s_{E}},0,1\\ m_{E},0,0\end{matrix}\right.\right).
\end{align}

Finally, substituting (\ref{sj1}), (\ref{sj2}) and (\ref{sj3}) into (\ref{defasc}), equation (\ref{sasc}) is obtained. Thus, the proof is completed.
\end{IEEEproof}

\subsection{Asymptotic Analysis} In what follows, we provide an asymptotic ASC analysis when $\overline{\gamma}_{D}\rightarrow\infty$ (i.e., $D$ is located close to $S$). According to \cite{Zhengdao}, the asymptotic behavior can be derived based on the behavior of the PDF of $\gamma_k$ around the origin. Thus, the PDF, given in (\ref{pdfr}) or (\ref{pdfr1}), and the CDF, given in (\ref{cdfr}), at high SNR regime can be written, respectively, as
\begin{align}\label{asypdf}
f_{k}(\gamma_{k})={a_{k}^{m_{k}}\over B(m_k,m_{s_{k}})}\gamma_{k}^{m_{k}-1}+\mathcal{O}\left((a_{k}\gamma_{k})^{m_{k}}\right),
\end{align}
and
\begin{align}\label{asycdf}
F_{k}(\gamma_{k})={a_{k}^{m_{k}}\Gamma(m_k+m_{s_{k}})\over\Gamma(m_k+1)\Gamma(m_{s_{k}})}\gamma_{k}^{m_{k}}+\mathcal{O}\left((a_{k}\gamma_{k})^{m_{k}}\right),
\end{align}
where $\mathcal{O}(\cdot)$ stands for higher-order terms.

\begin{corollary}[Asymptotic Average Secrecy Capacity] The asymptotic ASC for the system under consideration when the main channel and eavesdropper's channel experience i.n.i.d. Fisher-Snedecor $\mathcal{F}$ fading is given by (\ref{asymasc}).
\end{corollary}
\begin{IEEEproof}
The asymptotic ASC can be obtained via
\begin{align}\label{defSOP}
 \overline{C}_{s}^{\rm{ASC, asy}}  = {\mathcal{J}_{1}^{asy}} + {\mathcal{J}_{2}} - {\mathcal{J}_3},
 \end{align}
where ${\mathcal{J}_{2}}$ and ${\mathcal{J}_3}$ are defined in (\ref{sj2}) and (\ref{sj3}), respectively, and $\mathcal{J}_{1}^{asym}$ can be obtained using (\ref{cs1}) and (\ref{asypdf}) as follows
\begin{align}\label{j1asy}
&\mathcal{J}_{1}^{asym} = {a_{D}^{m_{D}}\over B(m_D,m_{s_{D}})\Gamma(m_{E})\Gamma(m_{s_{E}})}\int\limits_0^\infty \gamma_{D}^{m_{D}-1}{\ln \left({1 + {\gamma _D}} \right)}\cr &\qquad\qquad\qquad\qquad\qquad\times\mathrm{G}_{2,2}^{1,2}\left(a_E\gamma_D\left\vert \begin{matrix} 1-m_{s_{E}},1\\ m_{E},0\end{matrix}\right.\right)d{\gamma_D}.&
\end{align}
The latter integral can be solved, after representing the logarithmic function in terms of Meijer's G-function \cite[Eq. (3.4.6.5)]{a:pru} and using \cite[Eq. (2.24.1.1)]{a:pru}, as
\begin{align}\label{j1asysol}
&\mathcal{J}_{1}^{asy} = {a_{D}^{m_{D}}\over B(m_D,m_{s_{D}})\Gamma(m_{E})\Gamma(m_{s_{E}})}\cr& \qquad\qquad\qquad\times\mathrm{G}_{4,4}^{3,3}\left(a_E\left\vert \begin{matrix} 1-m_{s_{E}},1,-m_D,1-m_D\\ m_{E},-m_D,-m_D,0\end{matrix}\right.\right).\cr&
\end{align}
Finally, substituting (\ref{sj2}), (\ref{sj3}) and (\ref{j1asysol}) into (\ref{defSOP}), then (\ref{asymasc}) is obtained and thus the proof is completed.\end{IEEEproof}
\section{Results and Discussions}\label{sec6}
\begin{figure}[t] \centering
\hspace{-0.2in}
         \includegraphics[width=0.73\columnwidth]{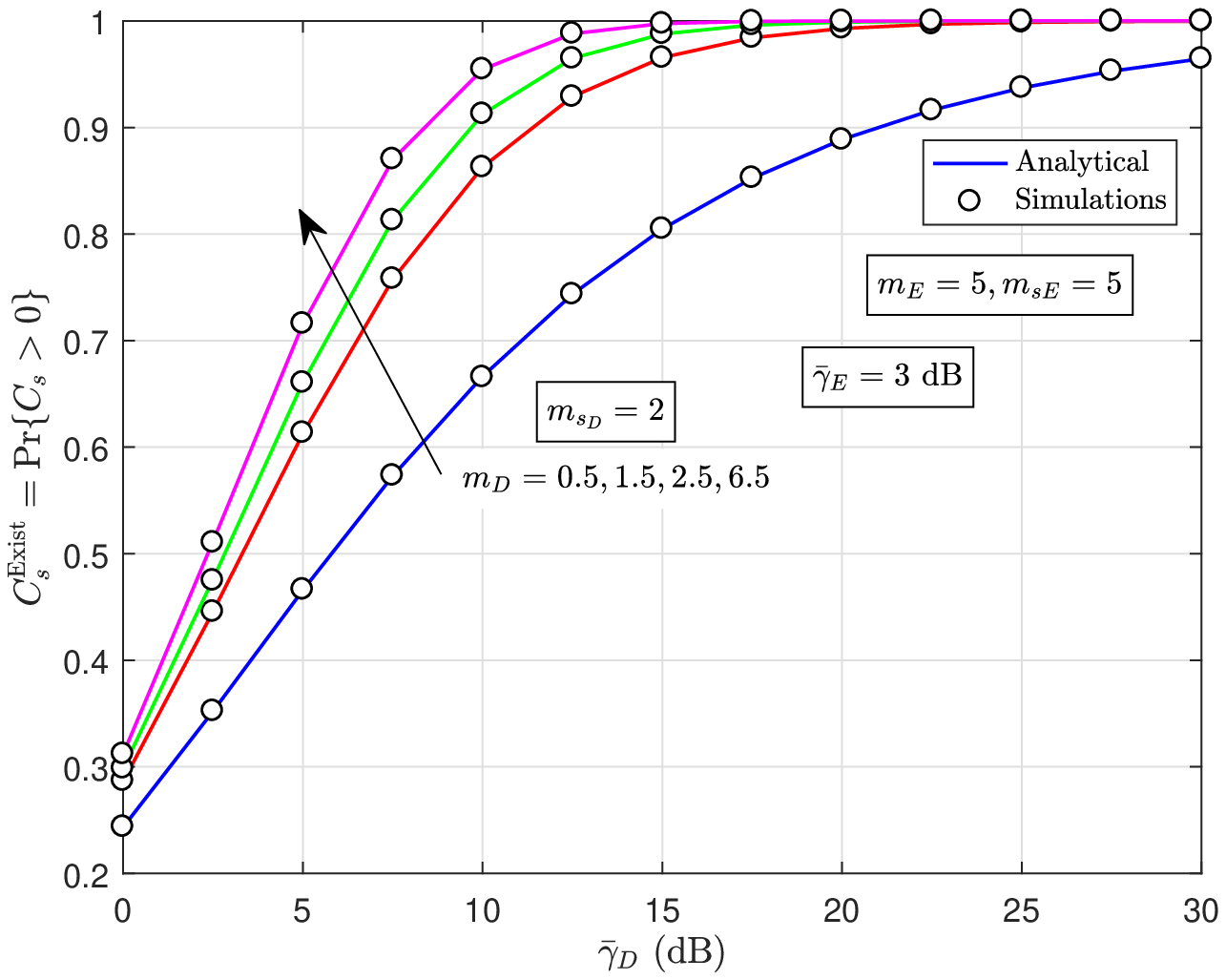}%
       \caption{The impact of $m_D$ on PNSC with respect to $\overline{\gamma}_{D}$.}\label{exmd}
\end{figure}
In this section, we present some analytical results to study the secrecy performance under different multipath fading and shadowing conditions. Throughout this section, Monte-Carlo simulation results are provided to validate the analysis. It can be seen from the figures that excellent agreement between the analytical results and Monte-Carlo simulation results is achieved throughout, thereby confirming the validity of the derived expressions.

\begin{figure}[t] \centering
\hspace{-0.2in}
         \includegraphics[width=0.73\columnwidth]{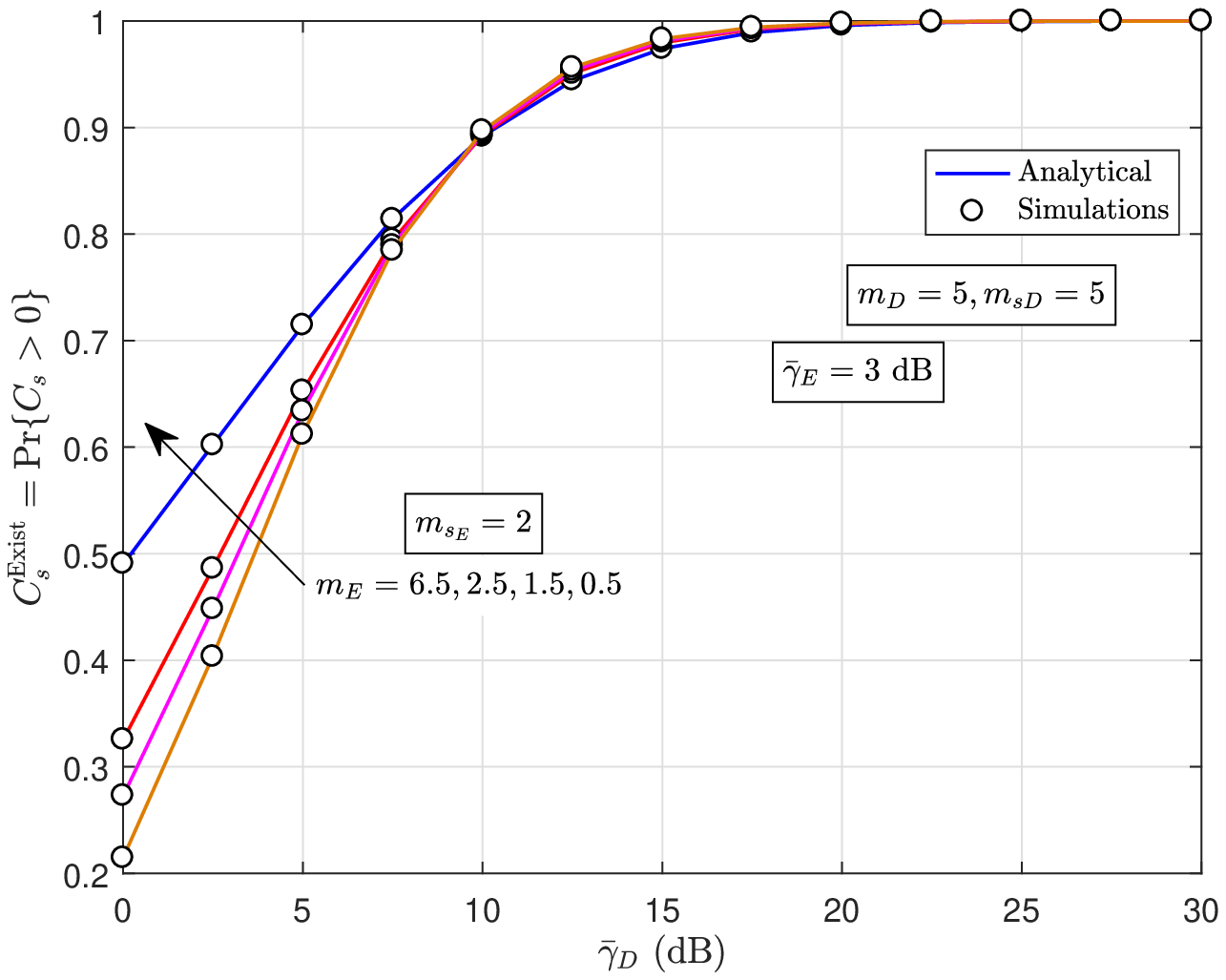}%
       \caption{The impact of $m_E$ on PNSC as a function of $\overline{\gamma}_{D}$.}\label{exme}
\end{figure}

In Fig. \ref{exmd} and Fig. \ref{exme}, we study the impact of the fading parameters $m_D$ and $m_E$ of the main channel and eavesdropper's channel, respectively, on the PNSC performance (i.e., $\Pr\{C_s>0\}$). Fig. \ref{exmd} shows that as $m_D$ increases (less fading severity in the legitimate link), the PNSC performance increases. In contrast to $m_D$, the results show that the PNSC performance increases as $m_E$ decreases (i.e. the multipath fading observed in the eavesdropper's channel increases). Conversely, as $m_E$ increases, the eavesdropper becomes more effective and the security of the system is degraded.

\begin{figure}[t] \centering
\hspace{-0.2in}
         \includegraphics[width=0.73\columnwidth]{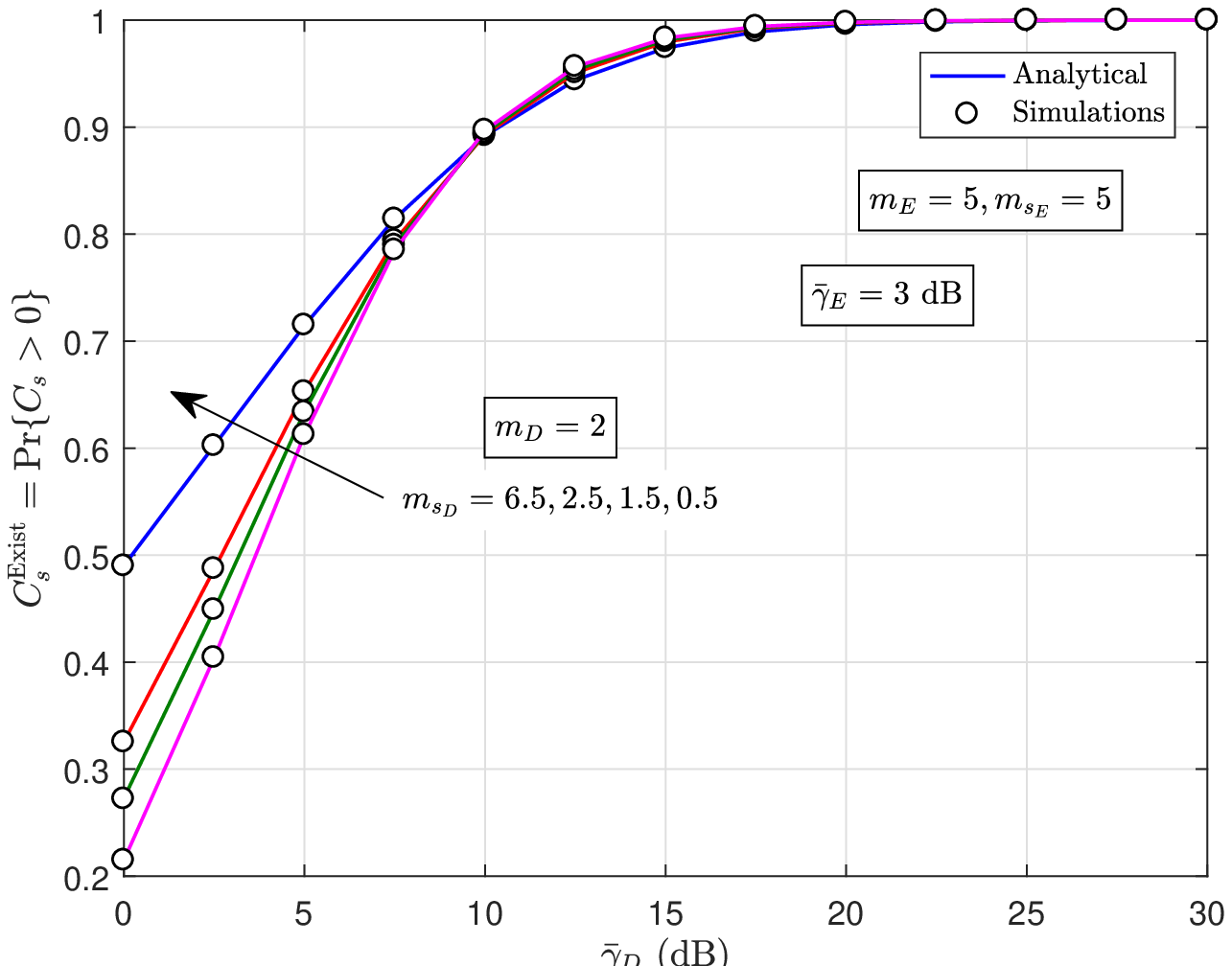}%
       \caption{The impact of $m_{s_{D}}$ on PNSC versus $\overline{\gamma}_{D}$.}\label{exmsd}
\end{figure}

\begin{figure}[b] \centering
\hspace{-0.2in}
         \includegraphics[width=0.73\columnwidth]{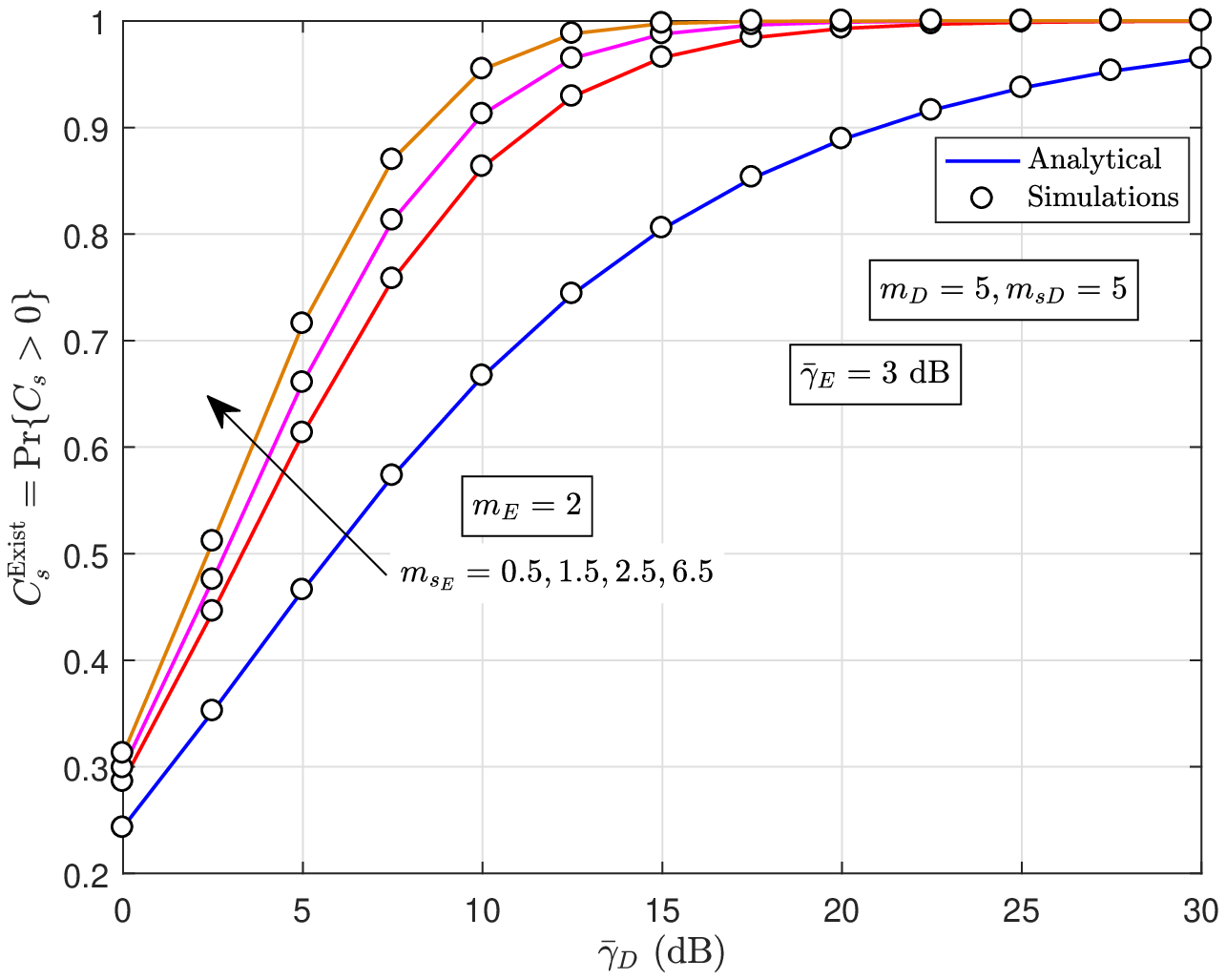}%
       \caption{The impact of $m_{s_{E}}$ on PNSC as a function of $\overline{\gamma}_{D}$.}\label{exmse}
\end{figure}

The influence of the shadowing parameters $m_{s_{D}}$ and $m_{s_{E}}$ of the main channel and eavesdropper's channel on the PNSC performance is illustrated in Fig. \ref{exmsd} and Fig. \ref{exmse}, respectively. It is clear that the PNSC improves as $m_{s_{D}}$ decreases, while the performance degrades as $m_{s_{E}}$ increases.

\begin{figure}[t] \centering
\hspace{-0.2in}
         \includegraphics[width=0.73\columnwidth]{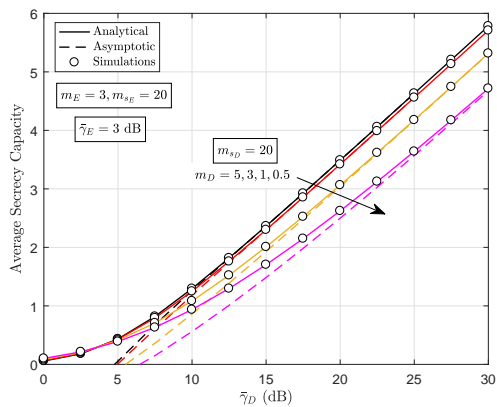}%
       \caption{The impact of $m_D$ on ASC with respect to $\overline{\gamma}_{D}$.}\label{ascmd}
\end{figure}

\begin{figure}[t] \centering
\hspace{-0.2in}
         \includegraphics[width=0.73\columnwidth]{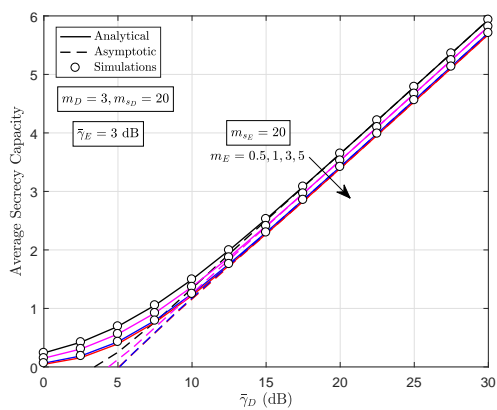}%
       \caption{The impact of $m_E$ on ASC as a function of $\overline{\gamma}_{D}$.}\label{ascme}
\end{figure}

Fig. \ref{ascmd} depicts the impact of the multipath fading parameter of the main channel (i.e., $m_D$) on the ASC performance.
As expected, the results demonstrate that the ASC performance improves as the average SNR $\bar{\gamma}_{D}$ increases. In addition, it can be seen that the ASC improves as the fading parameter $m_D$ increases. This is due to the fact that as $m_D$ increases, the number of multipath clusters arriving at the destination $D$ increases and thus, the received SNR increases.
On the other hand, the results of Fig. \ref{ascme} show that the ASC improves as the fading parameter of the eavesdropper's channel (i.e., $m_E$) decreases. Most striking though is the positive effect that $m_D$ can have on the ASC performance since it is more pronounced than that of $m_E$.

\section{Conclusions}\label{sec7}
In this paper, for the first time, we have investigated the secrecy performance of a single-hop wireless system in the presence of an eavesdropper over Fisher-Snedecor $\mathcal{F}$ fading channels. Specifically, exact closed-form expressions for the ASC and PNSC were derived. Additionally, an accurate asymptotic expression for the ASC when the destination is located close to the source was also derived. Monte-Carlo simulations were performed to validate the theoretical derivations, while a comprehensive numerical analysis has shown the impact of varying the intensities of multipath fading and shadowing on the secrecy performance. In particular, the results have demonstrated that increasing severity of multipath fading in the main channel and/or the shadowing of the eavesdropper's channel improve the secrecy performance.
\bibliographystyle{IEEEtran}
\bibliography{prodff}
\end{document}